\documentclass[fp,twocolumn]{jpsj3}
\usepackage{txfonts}

\usepackage{graphicx}
\usepackage{bm}

\title{Fractional quantum Hall effects in graphene on a h-BN substrate}

\author{Kouki Yonaga$^1$ and Naokazu Shibata$^2$}
\inst{$^1$AIST-TohokuU MathAM-OIL, Sendai 980-8577, Japan \\
$^2$Department of Physics, Tohoku University, Sendai 980-8578, Japan
} 

\abst{
Fractional quantum Hall (FQH) effects in graphene are studied because of their relativistic characteristics and the valley degree of freedom.
Recently FQH effects have been observed at various filling factors with graphene on a hexagonal boron nitride (h-BN) substrate.
However, it is known that h-BN creates the mass term in the Dirac Hamiltonian that acts as the effective model of graphene.
To understand recent experiments, we shall investigate many-body effects in the massive Dirac electron system.
In this paper, we study the mass-term effects on the FQH states of Dirac electrons by exact diagonalization.
We examine the ground state at filling factor 1/3 in the $n=\pm 1$ Landau level.
Without the mass term, the ground state in the Laughlin state featuring valley degeneracy and the lowest excitation is characterized by the valley unpolarized state (known as the valley skyrmion state).
Conversely, we find that the mass-term lifts the valley degeneracy due to the breaking of the inversion symmetry.
We also demonstrate that the valley unpolarized excitation is suppressed and that the fully or partially polarized state appears in the lowest excitation by increasing the mass term.
Finally, we discuss the stability of FQH states in the massive Dirac Hamiltonian in experimental situations.
We find that our numerical results are in agreement with previous experimental results.
}

\begin{document}
\maketitle

\section{\label{sec:intro}Introduction}
Graphene has attracted significant attention for several years due to its
interesting low-energy properties, as described by the massless Dirac Hamiltonian \cite{RevModPhys.81.109,RevModPhys.83.1193}.
The massless Dirac fermions in graphene originate from the symmetric AB sublattice of its hexagonal lattice structure. 
{Moreover, the equilateral triangular symmetry brings equivalence to the two Brillouin zone corners called $K$ and $K'$ points, where the character of the Dirac fermions appears in low-energy excitations.} 
The low-energy properties of graphene are described by the massless Dirac fermions in the valley $K$ and $K'$ whose valley degree of freedom is regarded as the pseudospin degree of freedom of Dirac fermions. A variety of interesting many-body phenomena are expected to be realized by Coulomb interaction between the Dirac fermions, and novel FQH states such as fully valley polarized FQH, valley unpolarized FQH states, and valley skyrmion excitations, have been theoretically proposed  \cite{PhysRevLett.97.126801,PhysRevB.74.235417,PhysRevB.77.235426,JPSJ.78.104708,PhysRevB.92.205120}.

Recently, these FQH states were studied on the substrate of h-BN \cite{Nanotech.5.722}, whose lattice constant is almost the same as that of graphene. The structural stability of graphene on h-BN is improved considerably, and high-quality results are obtained from the experiment. Several FQH states are clearly observed on the h-BN substrate, and the excitation gap energies of these many-body states are determined experimentally \cite{Nat.Phys.7.693,Nat.Commun.6.5838,Nat.Phys.8.550}. 
Although a h-BN substrate like this improves the stability of graphene, the potential energy difference arising from the sublattices of boron and nitrogen is expected to break the symmetry of the two sublattices of graphene. Because the breaking of the sublattice symmetry effectively modifies the interaction between the Dirac fermions in the valley $K$ and $K'$ due to the change in the wavefunction of the electrons, its effect on many-body states such as the FQH states is nontrivial although the valley degree of freedom and the sublattice degree of freedom are different.
The interaction between the valley and sublattice degrees of freedom affects the stability of the ground state, thus examining on the effect of the sublattice asymmetry is important for finding more stable FQH states. 

\begin{figure}[t]
 \centering
 \includegraphics[width=8cm,clip]{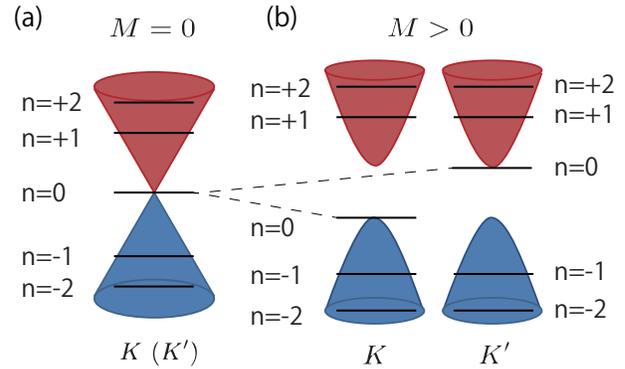}
 \caption{(Color online) Diagram of the energy spectrum and Landau levels; (a) and (b) show the massless ($M=0$) and the massive ($M>0$) cases, respectively. }
 \label{fig:spectrum}
\end{figure}

In this paper, we study the effect of the potential energy difference between the two sublattices of graphene, where the massless Dirac Hamiltonian is modified to produce sublattice dependent potentials $M$ and $-M$ in diagonal elements. The Hamiltonian, $H_{K(K')}$, around the valley, $K$($K'$), is
\begin{subequations}
\begin{align}
H_{K} &= p_x \sigma_x + p_y \sigma_y + M \sigma_z  \\
H_{K'} &= -p_x \sigma_x + p_y \sigma_y + M \sigma_z
\end{align}
\label{eq:Dirac_Hamiltonian}
\end{subequations}
where $p$ and $\sigma$ represent the momentum and the Pauli matrix acting on the AB sublattice components of the Dirac fermions, respectively\cite{Hunt1237240,PhysRevB.90.155406}. The first and second term, $\pm p_x \sigma_x + p_y \sigma_y$, in Eq.(\ref{eq:Dirac_Hamiltonian}) correspond to the  Hamiltonian of the massless Dirac fermions, whose low-energy eigenvalues form linear dispersion, as shown in Fig.\ref{fig:spectrum}(a). The final term, $M \sigma_z$, in Eq.(\ref{eq:Dirac_Hamiltonian}) generates a gap in the energy spectrum, with the resulting dispersion corresponding to the massive Dirac fermions (as shown in Fig.\ref{fig:spectrum}(b)). Therefore, $M\sigma_z$ is known as the mass term; Eq.(\ref{eq:Dirac_Hamiltonian}) is identical to the massive Dirac Hamiltonian when $M$ is finite. 
In this study, we presume that $M>0$ without any loss of generality. Note that $M$ in Eq.(\ref{eq:Dirac_Hamiltonian}) breaks both the sublattice symmetry and the spatial inversion symmetry.  

The importance of the mass term has already been demonstrated by the valley Hall effects in graphene \cite{PhysRevLett.99.236809,RevModPhys.82.1959}, and similar importance is also expected for the FQH effects. Indeed, DMRG studies by Shibata and Nomura demonstrate that the excited state of the massless Dirac Hamiltonian can be the valley unpolarized state (known as valley skyrmion) \cite{PhysRevB.77.235426,JPSJ.78.104708} and that the stability of the FQH states in graphene depends heavily on the valley degree of freedom. We can anticipate therefore that such a valley unpolarized state will be severely modified by the mass term whose effect on the FQH states remains unclarified. 

In this paper, we study the mass term dependence on the FQH states. We construct the effective Hamiltonian with the pseudopotentials, which depend on the strength of the mass term and the valley degree of freedom. By focusing on the FQH states at a filling factor, $\nu_{n=1}=1/3$, in the $n=\pm 1$ Landau level, we calculate the mass dependence of the ground state and the excitations by exact diagonalizations. We then analyze the mass dependence of the charge excitation gap and discuss the stability of FQH states in experimental situations.

\section{\label{sec:model}Model and Method}

\begin{figure}[t]
 \centering
 \includegraphics[width=7.5cm,clip]{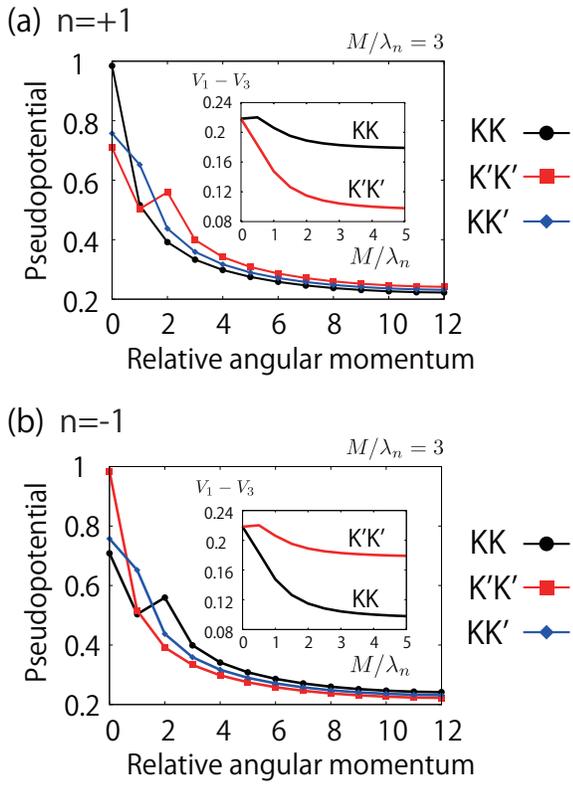}
 \caption{(Color online) Mass dependence of the pseudopotentials between electrons at each valley point. (a) and (b) show the results for $M/\lambda_n=3$ in $n=+1$ and $n=-1$, respectively. The insets in (a) and (b) show the short-range part of the pseudopotentials as a function of $M/\lambda_n$ where $\lambda_n$ is the single-particle energy at $M=0$. Here the red and black lines represent  $V_{1}^{KK}-V_{3}^{KK}$ and $V_{1}^{K'K'}-V_{3}^{K'K'}$, respectively.}
 \label{fig:PP}
\end{figure}

In this section, we derive the effective Hamiltonian of graphene on h-BN in a magnetic field. To simplify the calculation, we ignore the Landau level mixing and real spin degrees of freedom. The eigenenergy of Eq.(\ref{eq:Dirac_Hamiltonian}) in a magnetic field $B$ is 
\begin{equation}
E_{n} = {\rm sgn}(n) \sqrt{ \lambda_n^2 + M^2 }\ \ ( n=0,\pm 1, \pm 2, \cdots )
\end{equation}
where $\lambda_n = \hbar v_{\rm F} \sqrt{2 |n| }/l_B$ is the single-particle energy at $M = 0$, with $\hbar$, $v_{\rm F}$ and $l_B$ being the Planck constant, the Fermi velocity, and the magnetic length, respectively. 
The corresponding eigenstates in valley $K$ and $K'$ in $n \neq 0$ are
\begin{subequations}
\begin{align}
|n,m_s\rangle_{K} &=\frac{1}{ \sqrt{ 2E_{n}(E_{n} + M)} } \left[
\begin{array}{c}
 (E_{n} + M) ||n|-1,m_s \rangle \\
 \lambda_n ||n| ,m_s \rangle\\
 \end{array}\right] \\
&\ & \nonumber \\
|n,m_s \rangle_{K'} &= \frac{1}{ \sqrt{ 2E_{n}(E_{n} + M}) } \left[
\begin{array}{c}
 (E_{n} + M) ||n|,m_s \rangle \\
  \lambda_n ||n|-1,m_s \rangle \\
 \end{array} \right]
\label{eq:eigenstate}
\end{align}
\end{subequations}
and in $n=0$ are given as follows:
\begin{subequations}
\begin{align}
|0,m_s\rangle_{K} &=\left[
\begin{array}{c}
 0 \rangle \\
 |0 ,m_s \rangle\\
 \end{array}\right] \\
&\ & \nonumber \\
|0,m_s \rangle_{K'} &= \left[
\begin{array}{c}
 |0,m_s \rangle \\
 0 \rangle \\
 \end{array} \right]
\label{eq:eigenstate}
\end{align}
\end{subequations}
where $|n,m_s\rangle$ is the eigenstate of the conventional two-dimensional electron system in the $n$th Landau level and $m_s$ is the single-particle angular momentum in symmetric gauge \cite{QHE.Yoshioka}. The projected Coulomb interaction between the electrons in the $n$th Landau level is represented by Haldane's pseudopotential $V^n_m$ as follows:
\begin{equation}
H = \sum_{i<j}\sum_{\tau,\tau' = K,K'} \sum_{m} V_{m}^{n,\tau\tau'} P^{n}_m (r_{i},r_{j})
\label{eq:Hamiltonian}
\end{equation}
where $P^n_m$ is a projection operator onto states with a relative angular momentum, $m$, between the $i$th and the $j$th electron \cite{PhysRevLett.51.605}. The energy scale of the Coulomb interaction is given by $e^2/\epsilon l_B$ for the unit of energy (where $e$ is the elementary charge and $\epsilon$ is the dielectric constant). The explicit forms of the pseudopotentials are given by
\begin{subequations}
\begin{align}
&V^{n,KK}_{m} = f_1^n V_m^{n-1,n-1} + 2f_2^n V_m^{n,n-1} + f_3^n V^{n,n}_m  \\
&V^{n,K'K'}_{m} = f_3^n V_m^{n-1,n-1} + 2f_2^n V_m^{n,n-1} + f_1^n V^{n,n}_m  \\
&V^{n,KK'}_{m} = \nonumber \\
&\ \ \ \ \ \ \ (f_1^n + f_3^n)V_m^{n,n-1} + f_2^n (V_m^{n-1,n-1} + V_m^{n,n}) \\
&V^{n,K'K}_{m} = V^n_{m,KK'} 
\end{align}
\end{subequations}
where $V_m^{n,n'}$ is the conventional pseudopotentials for two-dimensional electrons, and the form factors, $f_n$, are defined as follows:
\begin{subequations}
\begin{align}
f_1^n &= \frac{(E_n +M)^2}{4E_n^2} \\
f_2^n &= \frac{\lambda_n^2}{4E_n^2} \\
f_3^n &= \frac{\lambda_n^4}{4E_n^2(E_n+M)^2},
\end{align}
\end{subequations}
respectively.

In order to study the many-body effects of Dirac electrons, we introduce the spherical geometry (known as Haldane sphere \cite{PhysRevLett.51.605}) and assume that all the electrons are bounded on its surface. 
The magnetic field is induced by a monopole of strength, $Q$, at the center of the sphere; thus its radius $R = l_B \sqrt{Q}$.
Each Landau level on the sphere are labeled by the angular momentum $l$ and the azimuthal angular momentum, $m_s=-l,-l+1,\cdots,l-1,l$.
Thus the $n$th Landau level has $(2l+1)$ degeneracy and satisfies $l = Q-n-1/2$ (where the factor of $1/2$ corresponds to spin connection for the Dirac fermions on the sphere \cite{Hasebe.10.1142}). 
We define the $n$th Landau level filling factor as $\nu_n = N_e/N_{\Phi}$ where $N_e$ is the number of the Dirac fermions and $N_\Phi$, which is equal to $2l$, corresponds to total flux through the surface of the sphere.

We derive the pseudopotential, $V^{n,n'}_m$, between the Dirac fermions on the sphere in our previous work \cite{PhysRevB.93.235122}. 
With the use of $V_m^{nn'}$, we study the effect of Coulomb interaction in the thermodynamic limit, $R \rightarrow \infty$. For Haldane sphere, the valley polarized Laughlin state at $\nu_n=1/3$ is only realized when the total flux $N_\Phi$ is given by the following \cite{PhysRevLett.54.237}:
\begin{equation}
N_{\Phi} = \nu_{n}^{-1} (N_e -1) = 2l.
\end{equation}
We define the total energy for the valley polarized state as
\begin{equation}
E_0(N_\Phi) = E_{\rm C}(N_\Phi) - \frac{ N_e^2 }{ 2R } 
\end{equation}
where the first term is the energy of Eq.(\ref{eq:Hamiltonian}) and the second term represents the effect of the neutralizing background and self-energy of the background. 
Because the stability of the Laughlin state is measured from the energy gap of the excitations, we calculate two different types of excitation energies: (1) valley polarized and (2) valley unpolarized or partially polarized excitation energies. 
We introduce the excitation energy, $E(N_{\Phi},\gamma)$ where $\gamma$ is the number of electrons in valley $K'$ ($K$) when the ground state is the fully polarized state in the valley $K$ ($K'$).
Thus, the excitation gaps are given by
\begin{equation}
\Delta_{c}^{\pm} (N_{\Phi},\gamma) = E(N_{\Phi}\pm 1,\gamma)-E_0(N_{\Phi})
\end{equation}
where $+$ ($-$) represents quasihole (quasiparticle) excitation. 
Thermal excitations in experiments are given by the sum of quasiparticles and quasiholes, and the minimum excitation gap is defined as bellow: 
\begin{equation}
\Delta_{c}(N_{\Phi},\gamma) = \Delta_{c}^{+} (N_{\Phi},\gamma) + \Delta_{c}^{-} (N_{\Phi},\gamma).
\end{equation}

In this paper, we calculate $E(N_\Phi)$ and $\Delta(N_\Phi,\gamma)$ by numerical exact diagonalization.
Moreover, to obtain the bulk properties, we define the energy and the charge gap in the thermodynamic limit, $N_\Phi \rightarrow \infty$, as follows:
\begin{eqnarray}
E_0 = \lim_{N_\Phi \rightarrow \infty } E_0(N_{\Phi}) \nonumber \\
\Delta_{c}(\gamma) = \lim_{N_\Phi \rightarrow \infty }\Delta_{c}(N_{\Phi},\gamma).
\label{eq:E_Delta}
\end{eqnarray}
When we calculate $E_0$ and $\Delta_c(\gamma)$, we sometimes have finite size-effects.
In order to get the good extrapolated values, we use a linear fitting with rescaled magnetic length $l_B' = \sqrt{N_{\Phi}\nu/N_e} l_B$ \cite{PhysRevB.66.075408}.

\section{\label{sec:result} Results}

\begin{figure}[t]
 \centering
 \includegraphics[width=5.5cm,clip]{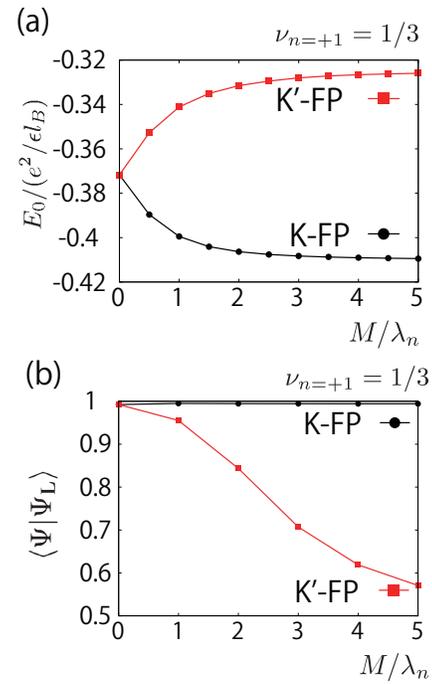}
 \caption{(Color online) (a) Mass dependence of the energy at the thermodynamic limit with unit $e^2/\epsilon l_B$; (b) shows the overlap between the valley polarized states and the Laughlin state. The black (red) line corresponds to the numerical result of the $K$ ($K'$) polarized state at $\nu_{n=+1}=1/3$.}
 \label{fig:energy}
\end{figure}

First, we observe the pseudopotentials, $V_m^{KK}$, $V_m^{K'K'}$ and $V_m^{KK'}$, which characterize the ground state of many-body wavefunctions. Fig.\ref{fig:PP} shows the relative angular momentum ($m$) dependence of the pseudopotentials at $M/\lambda_n=3$, where $\lambda_n$ is the single-particle energy of the massless Dirac Hamiltonian. In the $n=+1$ Landau level, the monotonic decrease of $V_m^{K'K'}$ in the massless limit ($M=0$) is modified by the mass term to form a local minimum structure at $m = 1$. 
The inset in Fig.\ref{fig:PP} (a) shows the mass dependence on the short-range part of the pseudopotentials ($V_1$, $V_3$): the red (blac) line corresponds to $V_{1}^{KK}-V_{3}^{KK}$ ($V_{1}^{K'K'}-V_{3}^{K'K'}$).
While $V_1^{KK}-V_3^{KK}$ retains large values, $V_1^{K'K'}-V_3^{K'K'}$ decreases as $M$ increases. Similar results are obtained in the $n=-1$ Landau level, where $V_m^{K'K'}$ decays monotonically  but $V_m^{KK}$ has a local minimum at $m = 1$ (as shown in Fig.\ref{fig:PP} (b)). Because there is symmetry between the $n = 1$ and $-1$ Landau levels, we focus on the ground state and the excitation in the $n = +1$ Landau level. 

Fig.\ref{fig:energy} (a) shows the mass dependence on the lowest energy at $\nu_{n=1}=1/3$. While the ground states of the valley polarized states in the valley $K$ and $K'$ are degenerate at $M=0$, this degeneracy is lifted by the mass term, $M$. Moreover, the energy difference increases as $M$ increases. 
The overlap between the polarized state in valley $K$($K'$) and the Laughlin state presented in Fig.\ref{fig:energy} (b) shows that the overlap keeps more than 99$\%$ of the overall $M$, although the overlap decreases rapidly as $M$ increases.
This means that the stability of the Laughlin state depends heavily on the short-range part in the pseudopotential around $m=1$ \cite{PhysRevLett.54.237}. 
As shown in Fig.\ref{fig:PP} (a), $V_m^{K'K'}$ has a local minimum at $m = 1$ and $V_1^{K'K'}-V_3^{K'K'}$ decreases with increasing $M$. 
Therefore, it is our understanding that the Laughlin state is not stabilized in valley $K'$ when $M$ increases, and we conclude that the ground state at $\nu_{n=+1}=1/3$ is the fully valley polarized Laughlin state in valley $K$.

\begin{figure}[t]
 \centering
 \includegraphics[width=8.5cm,clip]{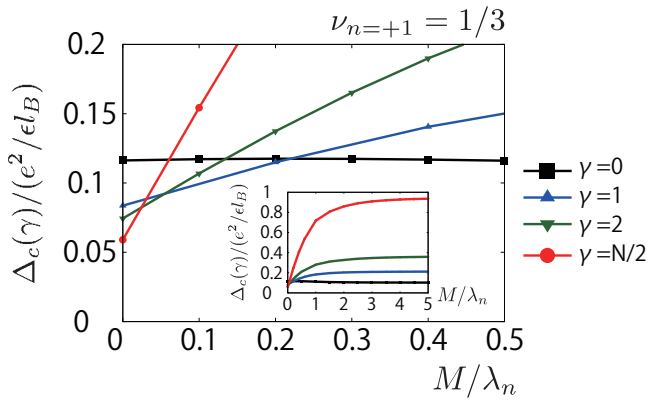}
 \caption{(Color online) Mass and $\gamma$ dependence of the charge gaps at the thermodynamic limit with unit $e^2/\epsilon l_B$. The inset shows the numerical results for $0 \le M/\lambda_n \le 5$. }
 \label{fig:gap}
\end{figure}

Next, we consider the excitations from the fully valley polarized Laughlin state. As explained in section \ref{sec:model}, the excited states can be in either the valley unpolarized or partially polarized state. To understand the characteristic features of the FQH states in graphene on h-BN, we investigate the mass dependence of the excitations with a variety of valley polarizations characterized by the difference between the number of electrons, $N_K$ and $N_{K'}$, in the valley $K$ and $K'$. The excitation energies obtained  from the fully polarized Laughlin state are shown as a function of $M$ in Fig.\ref{fig:gap}. 

{\renewcommand\arraystretch{1.5}
\begin{table}[t]
  \begin{tabular}{|c|c|c|} \hline
    B [T] & $\Delta_c^{M=0\ [\rm K]}(\gamma=N_e/2)$ [K] & $\Delta_c^{M=200\ [\rm K]}(\gamma=1)$ [K] \\ \hline
    10 &  20 & 42 \\
    15 &  25 & 53 \\
    20 &  29 & 64 \\ \hline
  \end{tabular}
 \caption{Magnetic field dependence of $\Delta_c^{M=200\ [\rm K]}(\gamma=1)$ and $\Delta_c^{M=0\ [\rm K]}(\gamma=N_e/2)$. Here, $\Delta_c^{M=200\ [\rm K]}(\gamma=1)$ and $\Delta_c^{M=0\ [\rm K]}(\gamma=N_e/2)$ correspond to the charge gap at $\nu_{n=\pm 1}=1/3$ in $M=200$ [K] and $M=0$ [K], respectively. We assume dielectric constant $\epsilon \sim 5$ and the single-particle energy $\lambda_n \sim 400\ \sqrt{B [{\rm T}]}$ [K].}
 \label{tb:gap}
\end{table}
}

In the massless limit, the sum of the quasiparticle and quasihole excitations is the lowest when the number of electrons in the valley $K$ and $K'$ are equal. This means that the lowest excitation is characterized by the valley unpolarized state, which is consistent with previous research \cite{PhysRevB.77.235426,JPSJ.78.104708}. However, the excitation energy to the unpolarized state increases as  the mass term is increased, and the polarization of the lowest charge gap changes such that $\gamma =N_e/2 \rightarrow \cdots \rightarrow 2 \rightarrow 1$. Therefore, the partially valley polarized excited states appear for  values of the intermediate $M$.
The fully valley polarized excitation ($\gamma=0$) shown in Fig.\ref{fig:gap} is almost independent of the mass term and becomes the lowest excitation energy for $M/\lambda_n > 0.2$. Therefore, the fully valley polarized excitation appears as the lowest excitation at the limit of a large mass.

Here, we consider why the $\gamma$ of the lowest excited state decreases as $M$ increases. 
As shown in Fig.\ref{fig:PP} (a), $V_m^{KK}$ and $V_m^{K'K'}$ display different $m$ dependences in $n = +1$ Landau level when $M\ne 0$. 
In particular, $V_1^{K'K'}-V_3^{K'K'}$  decreases as the mass term increases. 
Due to decreasing of the short-range part of the pseudopotential, the electrons at $K'$ point get close easily each other.
This means that the Coulomb energy of unpolarized excitation is enhanced by increasing $M$. 
Because a large $V_1^{KK}$ - $V_3^{KK}$ in the valley $K$ enhances the excitation energy from the beginning, the excitation energy of the unpolarized or partially polarized state becomes larger than that of the polarized state with the increase in the mass term. 

In the final part of this section, we compare our numerical results with the experimental date. The mass term is estimated to be about 50 $\sim$ 200 [K] in the experimental situation \cite{Nat.Commun.6.5838,Hunt1237240}.
This estimation indicates that only the partially valley polarized state is realized in experiments. In the following, we assume that the dielectric constant $\epsilon \sim 5$, the single-particle energy $\lambda_n \sim 400\ \sqrt{B [{\rm T}]}$ [K], and the mass term $M \sim 200$ [K], for reasons of simplicity \cite{Nat.Phys.7.693,Dielectric_Ando}.
When we consider $B = 10, 15,$ and $20$ [T], we can estimate $M/\lambda_n \sim 0.16, 0.13,$ and $0.11$, respectively. From these parameters, we find that the lowest excited state is characterized by $\gamma = 1$  (as shown in Fig.\ref{fig:gap}). 
We compare the charge gaps with the typical energy scale of disorder $\Gamma \sim  30$ [K] \cite{Nat.Phys.8.550}. 
The magnetic field dependence of $\Delta^{M=200\ [{\rm K}]}_c(\gamma = 1)$, which is the charge gap in $M=200\ [{\rm K}]$, is shown in Table \ref{tb:gap}. 
Moreover, we also show $\Delta_c^{M=0\ [\rm K]}(\gamma=N_e/2)$ which is the charge gap in the massless limit.
As illustrated in Table \ref{tb:gap}, $\Delta_c^{M=0\ [\rm K]}(\gamma=N_e/2)$ is smaller than $\Gamma$ in $B = 10,\ 15$, and $20$ [T]. 
If the excitation is given by the valley unpolarized state as expected in the previous works \cite{PhysRevB.77.235426,JPSJ.78.104708}, the FQH states can not be observed because the Laughlin state is smeared by disorder.
In contrast, $\Delta^{M=200\ [{\rm K}]}_c(\gamma = 1)$ is greater than the disorder energy scale all over $B$, indicating that the Laughlin state at $\nu_{n=1}=1/3$ is robust against disorder. 
This could be one reason why we observe FQH effects even when $B = 10$ [T]. Indeed, Amet $et\ al.$ observed clear FQH effects at $\nu = 7/3, 11/3$, and $13/3$  corresponding to $\nu_{n=1}=1/3$, which is consistent with our result.

\section{\label{sec:summary} Conclusion}
We have investigated the mass dependence of FQH states of the Dirac fermions using exact diagonalizations. It is shown that the pseudopotentials between the Dirac fermions are deformed through the mass term, and that the valley degeneracy is lifted. The ground state is characterized by the fully valley polarized Laughlin state. Although the valley unpolarized excitations are the lowest without the mass term, the partially valley polarized states reach the lowest excitations as the mass term increases. We find that the fully valley polarized excitations appear in the limit of large mass.

\section*{Acknowledgment}
This work was supported by {
JSPS KAKENHI Grant No. 26400344 and No. 16K05334.}

\nocite{*}


\end{document}